# Control of near-field radiative heat transfer based on anisotropic 2D materials

Lixin Ge,[1,2,a)] Yuping Cang,[1] Ke Gong,[1] Lihai Zhou,[1] Daqing Yu,[1] and Yongsong Luo[1]

[1]*School of Physics and Electronic Engineering, Xinyang Normal University, Xinyang 464000, China*

[2]*Division of Computer, Electrical and Mathematical Sciences and Engineering, King Abdullah University of Science and Technology (KAUST), Thuwal 23955-6900, Saudi Arabia*

In this work, we study the near-field radiative heat transfer between two suspended sheets of anisotropic 2D materials. It is found that the radiative heat transfer can be enhanced with orders-of-magnitude over the blackbody limit for nanoscale separation. The enhancement is attributed to the excitation of anisotropic and hyperbolic plasmonic modes. Meanwhile, a large thermal modulation effect, depending on the twisted angle of principal axes between the upper and bottom sheets of anisotropic 2D materials, is revealed. The near-field radiative heat transfer for different concentrations of electron is demonstrated and the role of hyperbolic plasmonic modes is analyzed. Our finding of radiative heat transfer between anisotropic 2D materials may find promising applications in thermal nano-devices, such as non-contact thermal modulators, thermal lithography, thermos-photovoltaics, etc.

Thermal radiation is an important physical phenomenon. Any object with temperature $T>0$K emits electromagnetic (EM) waves due to the fluctuating current generated from thermal motion of charge carriers. According to the Stefan-Boltzmann law, the radiative heat flux between two separated black bodies is given as $\langle S_{bb}\rangle = \frac{\pi^2 k_B^4}{60\hbar^3 c^2}(T_1^4 - T_2^4)$, where $k_B$ is the Boltzmann constant, $\hbar$ is the reduced Planck's constant, $c$ is the speed of light in vacuum, $T_1$ and $T_2$ is the high and low temperatures, respectively. For the Stefan-Boltzmann law, the separation distance $d$ is much larger than the thermal wavelength $\lambda_{th}=\hbar c/k_B T$ and only the propagation modes are taken into account in the process of radiative heat transfer. However, an additional contribution, i.e., evanescent waves, is dominant and should be considered when the separation distance $d$ is much smaller than $\lambda_{th}$ (about 10 μm at room temperature).[1-2] The evanescent waves near the surface can be surface plasmon polaritions (SPPs),[3] surface phonon polaritions (SPhPs),[3,4] or even frustrated modes from hyperbolic materials.[5-6] Due to large density of states of evanescent waves, near-field radiative heat transfer (NFRHT) can exceed the blackbody limit by several orders of

---

[a)] Author to whom correspondence should be addressed. Electronic mail: lixinge@hotmail.com

magnitude, which were reported in a number of configurations.[7-11] The control of NFRHT has many promising applications such as near-field imaging,[12] thermos-photovoltaics [13,14] heat-assisted magnetic recording,[15] etc.

The control of NFRHT based on isotropic two-dimensional (2D) materials such as graphene have been reported.[16-21] The frequency of SPPs of doped graphene lies in the mid-infrared, which guarantees a high-efficient excitation of SPPs at room temperature or above. Moreover, the property of large electrical tunability enables graphene as an excellent platform for active control of NFRHT.[16-21] Recently, anisotropic 2D materials, such as black phosphorus (BP),[22–25] rhenium disulfide (ReS$_2$) [26,27] and trichalcogenides,[28,29] have received enormous attentions. Particularly, monolayer or few-layer of black phosphorus, showing direct bandgap, high mobility and large current on/off ratios, is a promising 2D material for applications in optoelectronic devices.[30-32] Compared with the well-known graphene sheets, in-plane anisotropic is one of the most intriguing properties of anisotropic 2D materials. Under properly doping, SPPs can be supported in anisotropic 2D materials in the frequency of THz to infrared regime.[33] In addition to the features of sub-wavelength and electrical tunability, the SPPs in anisotropic 2D material are highly anisotropic. Moreover, natural hyperbolic SPPs can be supported in anisotropic 2D material under specific conditions.[33] These unique features of SPPs of anisotropic 2D materials are highly desired for the control of near-field thermal radiation.

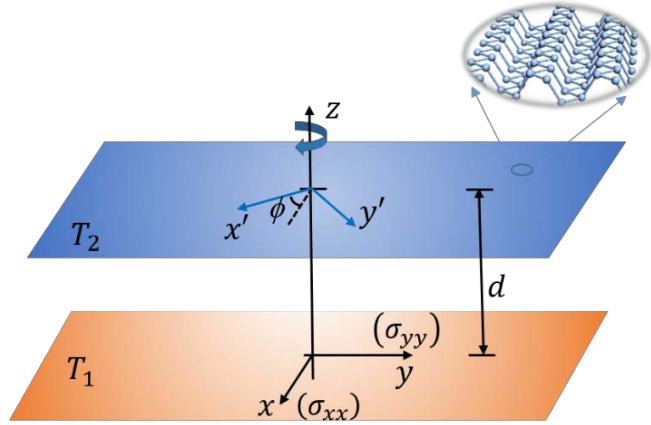

FIG. 1. Schematic diagram of near-field radiative heat transfer between two suspended sheets of anisotropic 2D material in the vacuum. The separation distance is *d*. The bottom sheet (temperature $T_1$) radiates EM waves to the upper sheet (temperature $T_2$). For the bottom sheet, two principal axes of conductivity tensor are parallel to the coordinate axes. There is a twisted angle (denoted by ϕ) of principal axes between the upper and bottom sheets, due to the rotation.

In this work, we investigated the NFRHT between two suspended sheets of anisotropic 2D materials. Based on the fluctuation-dissipation theory, the radiative heat flux between two separated sheets of anisotropic 2D materials is calculated. The enhancement can be orders-of-magnitude over the blackbody limit as the separation distance between 100 nm and 10 nm. It is revealed that the enhancement of NFRHT is not only related with the excitation of anisotropic plasmonic modes, but also with hyperbolic plasmonic modes. Moreover, the in-plane anisotropic properties of anisotropic 2D materials can provides a capability for thermal modulation based on rotation.

The system under study is depicted in Fig.1. Two parallel sheets of anisotropic 2D materials are suspended in vacuum and the spatial separation between them is $d$. The temperatures for the bottom and upper sheets are respectively $T_1$ and $T_2$. A twisted angle (denoted by $\phi$), i.e., relative crystalline orientation between the upper and bottom sheets of anisotropic 2D materials, is tunable through rotation. The optical conductivity for anisotropic 2D materials can be described by a tensor: [33]

$$\ddot{\sigma} = \begin{bmatrix} \sigma_{xx} & 0 \\ 0 & \sigma_{yy} \end{bmatrix}, \quad (1)$$

where

$$\sigma_{vv} = \frac{ie^2}{\omega + i\eta} \frac{n}{m_v} + s_v \left[ \Theta(\omega - \omega_v) + \frac{i}{\pi} \ln \left| \frac{\omega - \omega_v}{\omega + \omega_v} \right| \right], \quad (2)$$

$v=x, y$, where $e$ is the charge of an electron, $n$ is the concentration of electrons, $m_v$ is the effective mass of electron along the $v$ direction, $\eta$ corresponds to the relaxation time, $\Theta(\omega - \omega_v)$ is a step function defining the absorption due to the interband transition, $s_v$ represents the strength of interband transitions, and $\omega_v$ is the frequency of the onset of interband transitions for the $v$ component. The optical conductivity of a single sheet of anisotropic 2D is shown in Fig. 2(a) for a set of parameters. As the frequency is low (e.g., $\omega < 0.15$ eV/$\hbar$ for $n=5.0\times10^{13}$ cm$^{-2}$), the imaginary parts of $\sigma_{xx}$ and $\sigma_{yy}$ own the same sign, i.e., Im[$\sigma_{xx}$] Im[$\sigma_{yy}$]>0, which produces elliptic-like anisotropic plasmonic dispersion. At high frequency regime ($\omega>0.15$ eV/$\hbar$), it has Im[$\sigma_{xx}$] Im[$\sigma_{yy}$]<0, resulting in interesting hyperbolic plasmonic dispersion.[33]

The radiative heat flux between two sheets of anisotropic 2D materials can be calculated based on the fluctuation-dissipation theory, which is given as follows:[34]

$$\langle S \rangle = \frac{1}{(2\pi)^3} \int_0^\infty \left[ \int_{-\infty}^\infty \int_{-\infty}^\infty \xi(\omega, k_x, k_y) dk_x dk_y \right] [\Theta(\omega, T_1) - \Theta(\omega, T_2)] d\omega, \tag{3}$$

where $\Theta(\omega, T) = \hbar\omega/(\exp(\hbar\omega/k_B T)-1)$ is the average energy of a Planck's oscillator for angular frequency $\omega$ at temperature $T$, and $\xi(\omega, k_x, k_y)$ is the energy transmission coefficient, standing for the probability of photon tunneling. The energy transmission coefficient $\xi(\omega, k_x, k_y)$ are contributed from propagating and evanescent modes, expressed as follows: [34]

$$\xi(\omega, k_x, k_y) = \begin{cases} \mathrm{Tr}\left[ (\mathbf{I} - \mathbf{R}_2^\dagger \mathbf{R}_2) \mathbf{D}^{12} (\mathbf{I} - \mathbf{R}_1^\dagger \mathbf{R}_1) \mathbf{D}^{12\dagger} \right], & k_\parallel < \omega/c \\ \mathrm{Tr}\left[ (\mathbf{R}_2^\dagger - \mathbf{R}_2) \mathbf{D}^{12} (\mathbf{R}_1^\dagger - \mathbf{R}_1) \mathbf{D}^{12\dagger} \right] e^{-2|\gamma|d}, & k_\parallel > \omega/c \end{cases}, \tag{4}$$

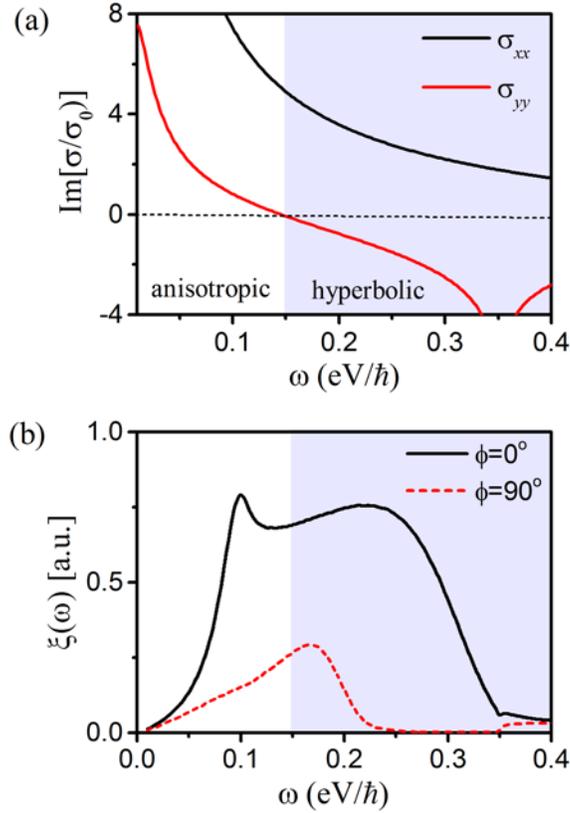

FIG. 2. (a) Imaginary part of optical conductivity of the anisotropic 2D material. Here we set $\omega_x=1.0$ (eV/$\hbar$), $\omega_y=0.35$ (eV/$\hbar$), $\eta=0.01$ (eV/$\hbar$), $m_x=0.2\ m_0$, $m_y=m_0$, $s_x=1.7 s_0$, $s_y=3.7 s_0$, $s_0=\sigma_0=e^2/4\hbar$ and $n=5.0\times 10^{13}$ cm$^{-2}$. $m_0$ is the mass of a free electron. (b) Spectral transmission function $\xi(\omega)$ for twisted angles $\phi=0°$ and $\phi=90°$. The separation distance $d=50$ nm. The shaded regime in (a) and (b) corresponds to frequency regime of hyperbolic SPPs.

where $k_\| = \sqrt{k_x^2 + k_y^2}$ and $\gamma = \sqrt{k_0^2 - k_\|^2}$ are respectively in-plan and vertical wavevectors, $k_0=\omega/c$ is the wavevector in vacuum, $\mathbf{D}^{12} = (\mathbf{I} - \mathbf{R}_1\mathbf{R}_2 e^{2i\gamma d})^{-1}$ and $\mathbf{R}_j$ ($j=1, 2$) is the 2×2 reflection matrix for the $j$-th sheet, having the form:

$$\mathbf{R}_j = \begin{pmatrix} r_j^{s,s} & r_j^{s,p} \\ r_j^{p,s} & r_j^{p,p} \end{pmatrix}, \tag{5}$$

where the superscripts $s$ and $p$ represent the polarizations of transverse electric (**TE**) and transverse magnetic (**TM**), respectively. The matrix element $r^{\alpha,\beta}$ ($\alpha, \beta = s, p$) represents the reflection coefficient for an incoming α-polarized plane wave turns out to be an outgoing β-polarized wave. Consider an incident plane wave with in-plane wavevector $\mathbf{k}_\| = k_x \vec{e}_x + k_y \vec{e}_y$, the reflection coefficients for single sheet of anisotropic 2D materials can be obtained as follows:[35,36]

$$\begin{aligned} r^{s,s} &= -\frac{2k_0^2 \bar{\sigma}'_{yy} + k_0\gamma(\bar{\sigma}'_{xx}\bar{\sigma}'_{yy} - \bar{\sigma}'_{xy}\bar{\sigma}'_{yx})}{4k_0\gamma + 2\gamma^2\bar{\sigma}'_{xx} + 2k_0^2\bar{\sigma}'_{yy} + k_0\gamma(\bar{\sigma}'_{xx}\bar{\sigma}'_{yy} - \bar{\sigma}'_{xy}\bar{\sigma}'_{yx})} \\ r^{p,p} &= -\frac{2\gamma^2 \bar{\sigma}'_{xx} + k_0\gamma(\bar{\sigma}'_{xx}\bar{\sigma}'_{yy} - \bar{\sigma}'_{xy}\bar{\sigma}'_{yx})}{4k_0\gamma + 2\gamma^2\bar{\sigma}'_{xx} + 2k_0^2\bar{\sigma}'_{yy} + k_0\gamma(\bar{\sigma}'_{xx}\bar{\sigma}'_{yy} - \bar{\sigma}'_{xy}\bar{\sigma}'_{yx})} \\ r^{s,p} &= r^{p,s} = \frac{2k_0\gamma\bar{\sigma}'_{yx}}{4k_0\gamma + 2\gamma^2\bar{\sigma}'_{xx} + 2k_0^2\bar{\sigma}'_{yy} + k_0\gamma(\bar{\sigma}'_{xx}\bar{\sigma}'_{yy} - \bar{\sigma}'_{xy}\bar{\sigma}'_{yx})} \end{aligned}, \tag{6}$$

where the elements of conductivity tensor are normalized by free-space impedance $\sqrt{\mu_0/\varepsilon_0}$, and the primes in superscripts stand for the rotation with respect to unit vector of $\mathbf{k}_\|$. For the bottom sheet, we have:

$$\bar{\sigma}' = \begin{bmatrix} \bar{\sigma}'_{xx} & \bar{\sigma}'_{xy} \\ \bar{\sigma}'_{yx} & \bar{\sigma}'_{yy} \end{bmatrix} = \begin{bmatrix} \bar{\sigma}_{xx}k_x^2/k_\|^2 + \bar{\sigma}_{yy}k_y^2/k_\|^2 & (\bar{\sigma}_{xx} - \bar{\sigma}_{yy})k_xk_y/k_\|^2 \\ (\bar{\sigma}_{xx} - \bar{\sigma}_{yy})k_xk_y/k_\|^2 & \bar{\sigma}_{xx}k_y^2/k_\|^2 + \bar{\sigma}_{yy}k_x^2/k_\|^2 \end{bmatrix}. \tag{7}$$

For the upper sheet, the anisotropic conductivity with respect to unit vector of $\mathbf{k}_\|$ is given by $\mathfrak{R}^T \bar{\sigma}' \mathfrak{R}$, due to an additional rotation angle ϕ, where $\mathfrak{R} = \begin{pmatrix} \cos\phi & -\sin\phi \\ \sin\phi & \cos\phi \end{pmatrix}$ and $\mathfrak{R}^T = \begin{pmatrix} \cos\phi & \sin\phi \\ -\sin\phi & \cos\phi \end{pmatrix}$. Interestingly, the off-diagonal elements of reflection matrix $r^{s,p} = r^{p,s} = 0$ for isotropic 2D materials because of $\bar{\sigma}'_{xy} = \bar{\sigma}'_{yx} = 0$. However, the off-diagonal terms of reflection matrix may nonzero for anisotropic 2D materials due to the anisotropic conductivity. Noted that the pole of reflection coefficients in Eq. (6) can produce the dispersion of SPPs of anisotropic 2D materials.[33]

According to above equations, the spectral transmission coefficient $\xi(\omega)$, i.e., the integral of $\xi(\omega, k_x, k_y)$ over the whole $k$-space, are shown in Fig. 2(b) for twisted angle $\phi=0^o$ and $\phi=90^o$. Clearly, the spectra, depending on the twisted angle, are consisted of frequency regimes for both anisotropic and hyperbolic SPPs. The spectral transmission coefficient $\xi(\omega)$ shows a broadband feature for $\phi=0^o$. However, the magnitude of $\xi(\omega)$ drop enormously and a strong suppressing of radiative spectrum is found at frequency regime (0.25 $eV$~0.35 $eV$) for $\phi=90^o$, which may find application in thermal spectral engineering. Noted that the hyperbolic regime for anisotropic 2D materials appears at high frequency, in contrast to that for patterned graphene ribbons.[18]

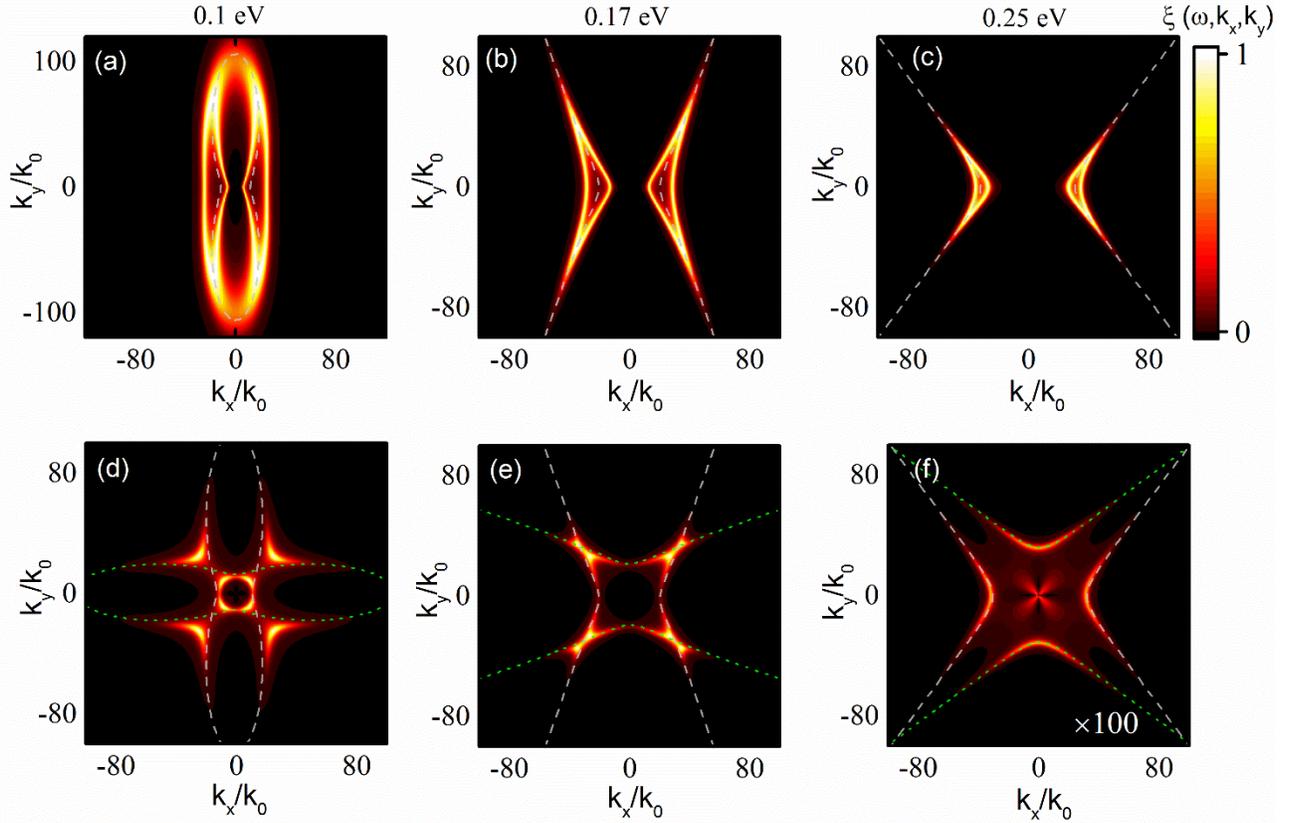

FIG. 3. Contours plots of energy transmission coefficient for different photon energies. (a) and (d) 0.1 eV; (b) and (e) 0.17 eV; (c) and (f) 0.25 eV. The twisted angle $\phi=0^o$ for upper panel, while it is $90^o$ for the bottom panel. The gray dashed curves represent the isofrequency of dispersion for single sheet of anisotropic 2D material, whereas the green dotted curves represent the corresponding isofrequency for a rotation of 90º. Noted that the magnitude of data in (f) has been magnified by 100 times.

To understand the broadband mechanism of $\xi(\omega)$ in Fig. 2(b), the contours plots of $\xi(\omega, k_x, k_y)$ at $k$-space are shown in Fig. 3 for three different photon energies. Firstly, we consider the case of $\phi=0^o$ (upper panel of Fig.3). For low photon energy $\hbar\omega=0.1$eV, Im$[\sigma_{xx}]$Im$[\sigma_{yy}]>0$ and an ellipse-like diagram can be observed, confirming the excitation of anisotropic SPPs. For high photon energies $\hbar\omega=0.17$ eV and 0.25 eV, Im$[\sigma_{xx}]$Im$[\sigma_{yy}]<0$ and hyperbolic configurations can be found in Fig. 3(b) and Fig.3(c). The broadband spectrum for $\phi=0^o$ is because of the coexistence of strong anisotropic and hyperbolic SPPs. Meanwhile, the large $k$ vector ($k_x$, $k_y >>k_0$) for both anisotropic and hyperbolic regimes guarantee a giant enhancement of radiative heat flux. For comparison, the corresponding isofrequency curves of SPPs dispersion of a single anisotropic 2D sheet are illustrated by the dashed gray lines. The strong coupling of SPPs for $\phi=0^o$ enables the original dispersion (dashed gray lines) splits into two bright branches in Figs. 3(a-c).

Now we consider the case for $\phi=90^o$ (bottom panel of Fig. 3). The contours of $\xi(\omega, k_x, k_y)$ show quite complicated diagrams, due to the mismatching coupling. For photon energies $\hbar\omega=0.1$ eV and 0.17 eV, the most effective coupling appears at $k_x=k_y$ as shown in Fig. 3(d) and (e). Nevertheless, the corresponding density of states becomes smaller compared with those of $\phi=0^o$. As the photon energy increases to $\hbar\omega=0.25$ eV, strong suppressing of radiative spectrum is confirmed by the contour plot of $\xi(\omega, k_x, k_y)$ in Fig. 3(f). Clearly, the magnitude of $\xi(\omega, k_x, k_y)$ is almost 100 times smaller than those of $\phi=0^o$, due to the weak coupling of SPPs. Correspondingly, the dispersion of the two separated sheets is almost the same as those of original dispersion of single sheet (dashed gray lines).

The radiative heat flux between two sheets of anisotropic 2D material, normalized by blackbody limit, are shown in Fig. 4(a). Remarkably, orders-of-magnitude enhancement can be found, especially at small separation for $\phi=0^o$ (e.g., about 2800 for $d$=10 nm in Fig. 4(a)). The enhancement become smaller as the twisted angle become $90^o$, indicating a thermal modulation effect based on twisted angle. In order to quantify the modulation effect, the ratios of radiative heat flux $<S(\phi)>/<S(\phi=0^o)>$ are given in Fig. 4(b). It can be seen that the thermal modulation is strong as $\phi$ increases from $0^o$ and it is almost unchanged as $\phi>45^o$. The ratio $<S(\phi=90^o)>/<S(\phi=0^o)>$ as a function of separation distance $d$ is also given in the inset of Fig. 4(b). The ratio can reduce to almost 25% for $d$=10 nm while it increases slightly as $d$ increases. It is worth mentioning that similar thermal modulation effect ware reported for two metallic/polar gratings[37] and two anisotropic hBN plates.[38] For anisotropic 2D materials, no pattern nanofabrication is needed, which is much simple for practical applications. Besides, the thermal modulation between anisotropic 2D materials can be enlarged further incorporated with tuning the doping level of concentration. The NFRHT as a function of concentration $n$ are shown in Fig. 4(d), which decreases greatly as $n$ varies from $1\times10^{13}$ cm$^{-2}$ to $10\times10^{13}$ cm$^{-2}$. Due to the exponentially decaying of Boltzman factor, low frequency of SPPs (corresponds to low density of concentration) is preferred to enhance NFRHT. The enhancement of NFRHT at $d=10$ nm can reach about $10^4$ times for n~$1\times10^{13}$ cm$^{-2}$, whereas it can drop to about $10^3$ times for $n$~$10\times10^{13}$ cm$^{-2}$.

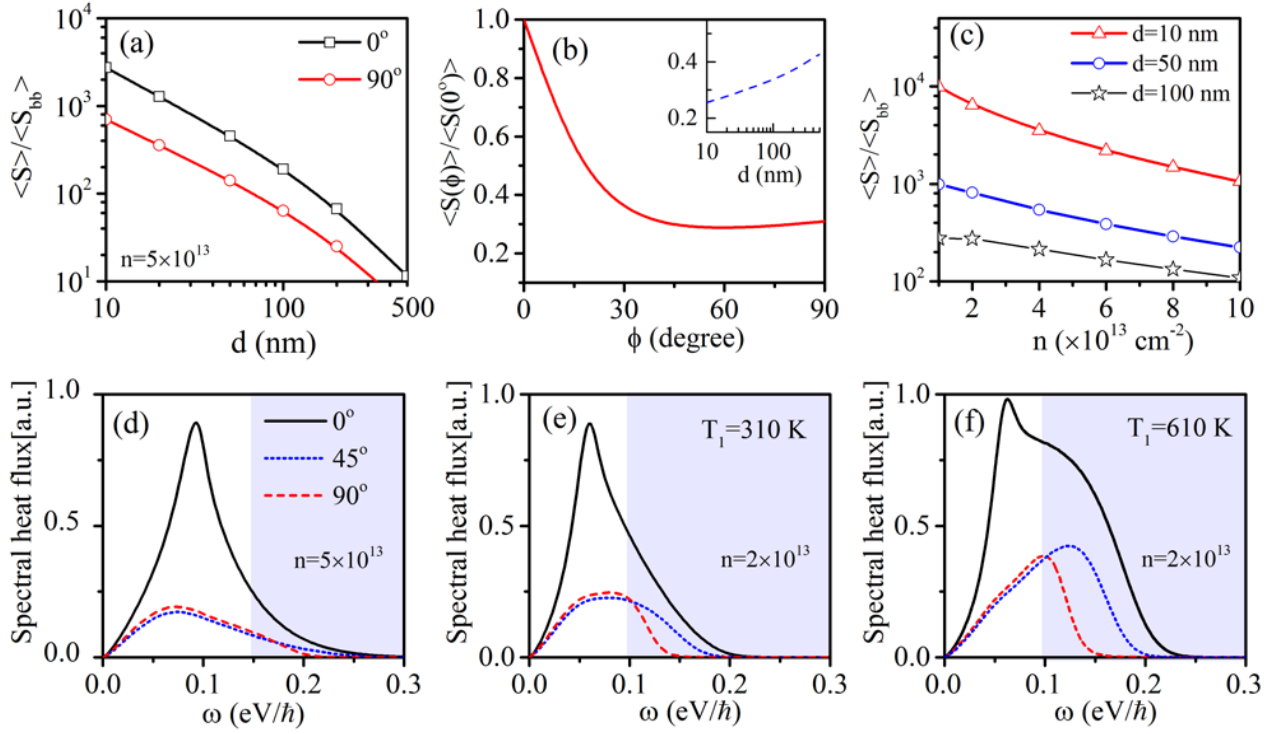

FIG. 4. (a) The radiative heat flux between two sheets of anisotropic 2D materials, normalized by blackbody limit. (b) Modulation effect as a function of twisted angle $\phi$, and we fixed $d=50$ nm. The inset shows the contrast of $<S(90^o)>/<S(0^o)>$ as a function of separation distance $d$. Here, the concentration of electron $n$~$5\times10^{13}$ cm$^{-2}$ is fixed in both (a) and (b). (c) The radiative heat flux as a function of $n$ for $\phi=0^o$. The spectral radiative heat flux for $n$ with (d) $5\times10^{13}$ cm$^{-2}$, (e) $2\times10^{13}$ cm$^{-2}$ and (f) $2\times10^{13}$ cm$^{-2}$. The high temperature $T_1=310$ K for (a)-(e), while $T_1=610$ K for (f), the low temperature $T_2=T_1-10$ K is fixed for different configurations. The shaded parts in (d-f) correspond to frequency regime of hyperbolic SPPs.

To understand NFRHT contributed from different SPPs modes. The spectral radiative heat flux for $n$~$5\times10^{13}$ cm$^{-2}$ and $n$~$2\times10^{13}$ cm$^{-2}$ are given respectively in Figs. 4(d) and 4(e) with $T_1=310$ K, $T_2=T_1-10$ K, and $d=50$ nm. For n~$5\times10^{13}$ cm$^{-2}$, anisotropic regime is dominant for NFRHT, while the contribution from hyperbolic regime occupies only a small part. As $n$ decreases to $2\times10^{13}$ cm$^{-2}$, the contribution from hyperbolic regime increases, but still smaller than those from anisotropic one.

Nevertheless, the radiative spectrum from hyperbolic SPPs can be enlarged as the temperature increases, due to the fact that the hyperbolic regime appears at relative high frequencies. Taking $T_1$=610 K as an example in Fig.4 (f), the radiative heat flux generated from hyperbolic SPPs are comparable to those from anisotropic one. Due to the coexistence of anisotropic and hyperbolic SPPs, the spectral radiative heat flux for anisotropic 2D materials is broadband, which can be 1 order broaden over than those of hBN.[38]

In summary, NFRHT between two suspended sheets of anisotropic 2D materials is investigated. The radiative heat flux can exceed the blackbody limit over $10^2$ to $10^4$ times for separation distance between 100 nm to 10 nm, stemming from the excitation of anisotropic and hyperbolic SPPs. The hyperbolic SPPs play an important role when the concentrations of electron are low and the temperatures are relative high. A large thermal modulation effect is demonstrated based on the twisted angle of principal axes between the upper and bottom sheets. Besides, active modulation of NFRHT can also be realized by tuning the concentrations of electron though external gating. Noted that the effective masses and band gaps of anisotropic 2D materials are tunable through external mechanical strain,[39, 40] or out-of-plane static electric fields,[41] which may provide alternative modulations of NFRHT without any rotation of the sheets. Our work may pave the way for applications of anisotropic 2D materials in non-contact thermal modulators, thermal lithography, thermos-photovoltaics, etc.

*Noted added.* After submission of this manuscript, we became aware of a preprint by Yong Zhang et al. (2018),[42] where the authors consider the NFRHT for suspended monolayer and multilayer black phosphorus. They investigated thoroughly and came to some similar conclusions. It should be pointed out that the reflection coefficients here are calculated by a different method, which is much simpler than that adopted by Yong Zhang et al.[42] The conductivity model we adopted in eqs. (1-2) can be applied not only to black phosphorus but also other anisotropic 2D materials. Besides, the role of hyperbolic plasmonic modes in NFRHT is analyzed separately in our work.

## ACKNOWLEDGMENTS

This work is supported by the National Natural Science Foundation of China (Grant No. 11747100), and the Innovation Scientists and Technicians Troop Construction Projects of Henan Province. The research of L.X. Ge was further supported by Nanhu Scholars Program for Young Scholars of XYNU.